\documentclass[12pt,a4paper,final]{iopart}

\usepackage{iopams}
\usepackage{graphicx}
\usepackage{color}
\usepackage[breaklinks=true,colorlinks=true,linkcolor=blue,urlcolor=blue,citecolor=blue]{hyperref}

\begin{document}

\title[]{Asymptotic non-flatness of an effective black hole model based on loop quantum gravity}

\author{Mariam Bouhmadi-L\'{o}pez$^{1,2}$, Suddhasattwa Brahma$^{3,4}$, Che-Yu Chen$^{5,6}$, Pisin Chen$^{5,6,7}$, and Dong-han Yeom$^{8,9}$}
\address{$^1$Department of Theoretical Physics, University of the Basque Country UPV/EHU, P.O. Box 644, 48080 Bilbao, Spain}
\address{$^2$IKERBASQUE, Basque Foundation for Science, 48011, Bilbao, Spain}
\address{$^3$Asia Pacific Center for Theoretical Physics, Pohang 37673, Republic of Korea}
\address{$^4$Department of Physics, McGill University, Montr\'eal, QC H3A 2T8, Canada}
\address{$^5$Department of Physics and Center for Theoretical Sciences, National Taiwan University, Taipei, Taiwan 10617}
\address{$^6$LeCosPA, National Taiwan University, Taipei, Taiwan 10617}
\address{$^7$Kavli Institute for Particle Astrophysics and Cosmology, SLAC National Accelerator Laboratory, Stanford University, Stanford, CA 94305, USA}
\address{$^8$Department of Physics Education, Pusan National University, Busan 46241, Republic of Korea}
\address{$^9$Research Center for Dielectric and Advanced Matter Physics, Pusan National University,
Busan 46241, Republic of Korea}

\ead{mariam.bouhmadi@ehu.eus, suddhasattwa.brahma@gmail.com, b97202056@gmail.com, pisinchen@phys.ntu.edu.tw, innocent.yeom@gmail.com}

\begin{abstract}
For any non-rotating effective quantum (uncharged) black hole model to be viable, its asymptotic structure of spacetime should reduce to that of a Schwarzschild black hole. After examining the asymptotic structure of quantum black holes proposed in \cite{Ashtekar:2018cay,Ashtekar:2018lag}, we find that the solution is actually not asymptotically flat. Departure from asymptotically flat spacetime is a direct consequence of the quantum parameters included in the model. Moreover, since the property of asymptotic flatness needs to be explicitly invoked, for instance, in relating the black hole and the white hole masses, it thereby raises a serious concern about the consistency of this model in view of the classical limit.{\footnote{Very recently, the authors of \cite{Ashtekar:2020ckv} have investigated independently the asymptotic spacetime structure of the model of \cite{Ashtekar:2018cay,Ashtekar:2018lag} in detail. They found that the fall-off of the curvature is slower than that of Schwarzschild spacetime. This pathology had, in fact, already been pointed out in the v2 version on arXiv of this paper (arXiv:1902.07874v2), and will be explicitly shown in this version.}}
\end{abstract}

\vspace{2pc}

\section{Introduction}
Even though general relativity (GR) has proven to be extremely accurate in describing our universe, it predicts the existence of spacetime singularities such as that inside the black hole. It is believed that some sort of quantum effects should be taken into account in order to naturally resolve the singularity problem. However, so far there is no consensus on how a fundamental theory encapsulating gravity and quantum effects should be built and it is still currently an intensive research arena. One of the mainstream approaches in this direction is loop quantum gravity (LQG). A popular approach to address the black hole singularity problem based on LQG has been to consider effective models that include some of the non-perturbative quantum effects of the theory. This is a first-step in realizing the physical implications of the putative quantum geometry, which is predicted in LQG, for black hole spacetimes.\footnote{It should be emphasized that there are other non-perturbative approaches towards the formulation of quantum gravity, such as those through the canonical quantum gravity approach \cite{qgkiefer}, string theory \cite{Strominger:1996sh} and the Euclidean path-integral approach \cite{Chen:2018aij}.} It is rather intriguing that the property of quantum geometry encoded in such a non-perturbative quantum theory of gravity can be captured by effective models, in which the quantum spacetime properties can usually be scrutinized systematically, sometimes analytic solutions are even attainable. More surprisingly, the singularity-resolution is a common feature shared by these effective models based on LQG.  The success of these approaches has strongly motivated further investigations following this line in LQG.

Typical effective models, which robustly examine the issue of singularity resolution in black holes due to the unique quantum effects implied by the theory, include the so-called holonomy modifications derived from LQG \cite{Ashtekar:1995zh}. The combination of having a minimum area-gap (at Planck scales) along with the necessity of working with holonomies (or parallel transport of connections), instead of the connections themselves, in the quantum theory gives rise to such corrections in LQG \cite{Ashtekar:1996eg,Ashtekar:2003hd}. Holonomy modifications in effective models are usually implemented via the so-called polymerization technique, which replaces the conjugate momenta $p$ in the phase space with their polymerized counterparts $\sin(\lambda p)/\lambda$, where $\lambda$ stands for a quantum parameter related to the area-gap. (The latter trigonometric quantities can be regarded as matrix element of holonoies.) Due to this, classical dynamics could be significantly modified at large curvature scales in these effective models, whereas the models are expected to recover the classical limit by sending $\lambda$ to zero.  

In previous studies of resolving black hole singularities within the LQG framework, there have been two main classes of investigations. The first one is the $\mu_0$-type schemes \cite{Ashtekar:2005qt,Modesto:2005zm,Campiglia:2007pr,Modesto:2008im,Gambini:2013ooa,Bojowald:2016itl}. In this approach, the quantum parameters are assumed to be constant on the entire phase space. However, some problems may appear in this setup, such as that the final results may depend on the fiducial  structures, which are initially introduced to construct the classical phase space. In addition, significant quantum effects may emerge at the low curvature regime rendering these models unphysical. On the other hand, in the so-called $\bar\mu$-type schemes \cite{Bohmer:2007wi,Chiou:2008nm,Chiou:2008eg,Joe:2014tca}, the quantum parameters are allowed to be functions of phase space variables. In this approach, there could still be huge quantum modifications on the event horizon. See also \cite{Bodendorfer:2019cyv} for the construction of effective models based on a new classical phase space description and \cite{Bojowald:2018xxu,BenAchour:2018khr} for effective black hole models including deformed covariance in LQG.

Recently, the authors in \cite{Ashtekar:2018cay,Ashtekar:2018lag} proposed a generalized version of the $\mu_0$-type schemes, which we shall refer to as the AOS model. In this effective model, the quantum parameters are chosen such that they are functions of the effective Hamiltonian itself. In this sense, the quantum parameters turn out to be Dirac observables, that is, constant only along the dynamical trajectories. This model is quite interesting because it not only solves the singularity problem, but also does not suffer from the aforementioned drawbacks, such as the appearance of quantum effects at low curvature regimes, dependence on fiducial structures, and mass amplifications from crossing through the transition surfaces. In \cite{Ashtekar:2018cay,Ashtekar:2018lag}, the authors also extended the effective models to its exterior spacetime, intended to construct an effective picture which smoothly connects the interior and the exterior regions. It is exciting that the AOS model, based on the novel formulation regarding the definition of its quantum parameters in terms of phase space variables, is able to address so many long-lasting issues in the LQG community. Therefore, the AOS model definitely deserves deeper and further analysis. Under more careful scrutiny, however, it was shown that there exist subtle complications in the AOS model \cite{Bodendorfer:2019xbp}.

For an effective black hole constructed in LQG, it is expected that quantum effects are significant only inside the horizon. This can be treated as a necessary requirement in order for the model to be viable. When moving away from the center of black hole, the quantum effects become negligible and the theory reduces to GR in vacuum (for an isolated black hole). More precisely, the black hole spacetime should reduce to the Schwarzschild spacetime (if the spacetime is static and spherically symmetric), when sufficiently outside the event horizon, which is asymptotically flat. This implies that the effective black hole should also be asymptotically flat. The physical meaning of asymptotic flatness is that the gravitational field approaches zero at large distances from an isolated black hole. We regard this property as an important requirement for an isolated black hole model since local objects should not change the asymptotic structure of spacetime. Therefore, in absence of matter field, any effective quantum-corrected black hole which is not asymptotically flat should not be physically viable.

The main objective of this letter is to point out that the exterior spacetime proposed in \cite{Ashtekar:2018cay,Ashtekar:2018lag} has a serious flaw. In fact, the singularity-resolution in this model comes at a very heavy cost of destroying a global (or, asymptotic) property of the spacetime. Even though the interior spacetime recovers the Schwarzschild metric near the event horizon and can be smoothly connected to the exterior, the full spacetime itself is \textit{not asymptotically flat}. In the asymptotic region, this effective model does not reduce to the Schwarzschild spacetime because of the presence of quantum parameters. Such deviations in the asymptotic region raises serious questions on the viability of the model and its claim that quantum effects exist only deep inside the black hole horizon. Indeed, as discussed in \cite{Bojowald:2019dry}, these shortcomings might be pointing towards a deeper malaise.

\section{The AOS quantum black hole}
As mentioned earlier, the AOS quantum black hole proposed in \cite{Ashtekar:2018cay,Ashtekar:2018lag} is an effective description of Schwarzschild spacetimes in the context of LQG. Using the fact that the interior of the Schwarzschild black hole is isometric to the Kantowski-Sachs spacetime, the interior region is foliated by spacelike homogeneous surfaces with coordinates $x$, $\theta$, and $\phi$. Within this setup, the components of the SU(2) connections $A_a^i$ can be described by two variables $b$ and $c$. Their conjugate momenta $E_i^a$, on the other hand, are described by the variables $p_b$ and $p_c$. In this regard, ($b$, $p_b$); ($c$, $p_c$) are two canonically conjugate pairs \cite{Ashtekar:2018cay,Ashtekar:2018lag}. With an appropriate choice of the time coordinate $T$ and its corresponding lapse function $N$, the interior region can be described by the following metric
\begin{equation}
ds^2=-N^2dT^2+\frac{p_b^2}{\left|p_c\right|L_0^2}dx^2+\left|p_c\right|d\Omega^2\,,\label{metricinq}
\end{equation}
where $L_0$ is the size of the fiducial cell on the 3-manifold and the physically viable solution should not depend on it. The Poisson brackets between the Ashtekar-Barbero connection and the triad variables lead to $[c,p_c]=2G\gamma$ and $[b,p_b]=G\gamma$, where $\gamma$ is the Immirzi parameter. The  Schwarzschild form of the interior region can be recovered by choosing the lapse function $N_{cl}^2=\gamma^2\left|p_c\right|/b^2$ and using the Hamiltonian constraint
\begin{equation}
H_{cl}\left[N_{cl}\right]=-\frac{1}{2G\gamma}\left[2cp_c+\left(b+\frac{\gamma^2}{b}\right)p_b\right]\,.
\end{equation}
In this regard, the mass of the black hole $m\equiv cp_c/L_0\gamma$ can be proven to be a Dirac observable, i.e., a constant of motion along the dynamical trajectory.

The effective modifications to the above classical lapse function and Hamiltonian in \cite{Ashtekar:2018cay,Ashtekar:2018lag} are introduced through the two quantum (polymerization) parameters $\delta_b$ and $\delta_c$. These parameters are assumed to be Dirac observables as well, in the sense that they should commute with the effective Hamiltonian. The lapse function and the effective Hamiltonian for the interior spacetime read
\begin{eqnarray}
N^2=\frac{\gamma^2p_c\delta_b^2}{\sin^2\left(\delta_bb\right)}\,,\nonumber\\
H_{eff}[N]=-\frac{1}{2G\gamma}\left[\frac{2\sin\left(\delta_cc\right)}{\delta_c}p_c+\left(\frac{\sin\left(\delta_bb\right)}{\delta_b}+\frac{\gamma^2\delta_b}{\sin\left(\delta_bb\right)}\right)p_b\right]\,.
\end{eqnarray}
When $\delta_b\rightarrow 0$ and $\delta_c\rightarrow 0$, the lapse function and the Hamiltonian recover their classical limits. After solving the equations of motion generated by the effective Hamiltonian, it can be shown that the quantum corrected black hole is free of spacetime singularities. Indeed, the interior singularity is replaced with a spacelike transition surface separating a trapped and an anti-trapped regions. In addition, at the transition surface the curvature acquires its maximum scale which is independent of the black hole mass for a macroscopic black hole, a requirement necessary for the validity of the effective description \cite{Ashtekar:2018cay,Ashtekar:2018lag}. 

In addition to the interior solution, the authors of \cite{Ashtekar:2018cay,Ashtekar:2018lag} extended the investigations to the exterior region by foliating it with timelike homogeneous surfaces (e.g. constant $r$ surfaces in Schwarzschild coordinates) such that the Ashtekar-Barbero connections ($\tilde{b}$ and $\tilde{c}$) and their canonical conjugate momenta ($\tilde{p}_b$ and $\tilde{p}_c$) take values in $\textrm{SU}(1,1)$ rather than in $\textrm{SU}(2)$. The phase space variables describing the exterior dynamics are related to the interior via the substitution
\begin{equation}
b\rightarrow i\tilde{b}\,,\qquad p_b\rightarrow i\tilde{p}_b\,,\qquad c\rightarrow\tilde{c}\,,\qquad p_c\rightarrow\tilde{p}_c\,.
\end{equation}
The exterior metric reads
\begin{equation}
d\tilde{s}^2=-\frac{\tilde{p}_b^2}{\tilde{p}_cL_0^2}dx^2-\tilde{N}^2dT^2+\tilde{p}_cd\Omega^2\,.\label{metricoutq}
\end{equation}
This method of deriving the exterior solution does work for classical solutions, regaining the well-known form of the Schwarzschild exterior. Furthermore, the interior and exterior regions can be smoothly connected at the event horizon. Later, we will show that the exterior solution of the effective spacetime derived using this approach is not asymptotically flat. The presence of the quantum parameters unavoidably alters the asymptotic behavior of the exterior region, jeopardizing the validity of the whole solution.

By choosing a proper lapse function $\tilde{N}^2=-\gamma^2 \tilde{p}_c\delta_b^2/\sinh^2{(\delta_b\tilde{b})}$, one obtains an effective Hamiltonian for the exterior spacetime from which the equations of motion can be solved as follows \cite{Ashtekar:2018cay,Ashtekar:2018lag}
\begin{eqnarray}
\tan\left(\frac{\delta_c\tilde{c}\left(T\right)}{2}\right)=\frac{\gamma L_0\delta_c}{8m}e^{-2T}\,,\\
\tilde{p}_c\left(T\right)=4m^2\left(e^{2T}+\frac{\gamma^2L_0^2\delta_c^2}{64m^2}e^{-2T}\right)\,,\label{34}\\
\cosh\left(\delta_b\tilde{b}\left(T\right)\right)=b_0\tanh\left(\frac{1}{2}\left(b_0T+2\tanh^{-1}\left(\frac{1}{b_0}\right)\right)\right)\,,\label{35}\\
\tilde{p}_b\left(T\right)=-2m\gamma L_0\frac{\sinh\left(\delta_b\tilde{b}\left(T\right)\right)}{\delta_b}\frac{1}{\gamma^2-\frac{\sinh^2\left(\delta_b\tilde{b}\left(T\right)\right)}{\delta_b^2}}\,.
\end{eqnarray}
 The two quantum parameters $\delta_b$ and $\delta_c$ share the same values with their interior counterparts and similarly they are assumed to be constant along the dynamical trajectories  \cite{Ashtekar:2018cay,Ashtekar:2018lag}. Here, $b_0\equiv\sqrt{1+\gamma^2\delta_b^2}$ has been introduced for the sake of abbreviation. As expected, $m =p_c\sin\left(\delta_cc\right)/\left(\gamma L_0\delta_c\right)$, a Dirac observable, denotes the mass of the black hole.

\section{Asymptotic structure}
 
\begin{figure}
\center
	\includegraphics[scale=0.7]{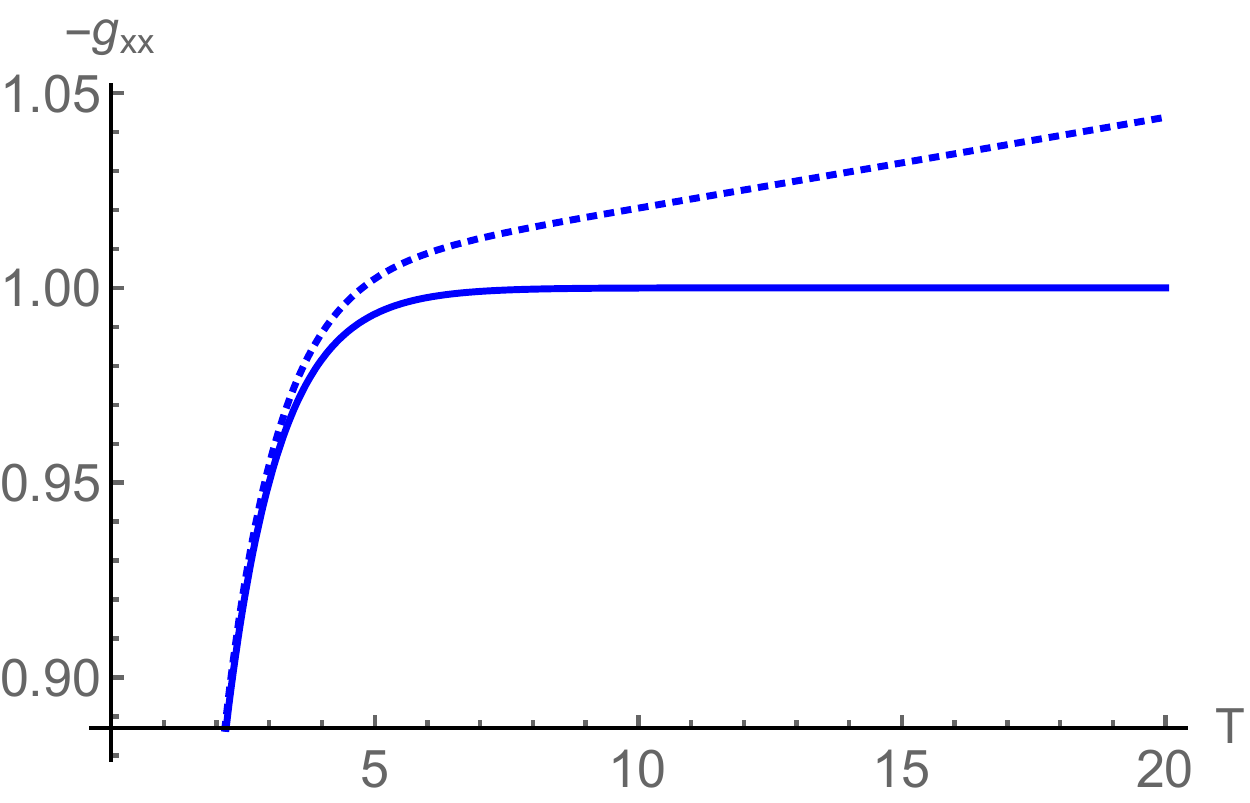}
	\caption{\label{f2} The metric function $-g_{xx}$ of the exterior AOS spacetime (dotted) is shown as a function of $T$. The solid curve corresponds to the Schwarzschild metric. When approaching the event horizon, the AOS black hole recovers the Schwarzschild solution, but they deviate significantly from each other when moving away from the black hole.}
\end{figure}

To study the asymptotic structure of the AOS metric (\ref{metricoutq}), we shall transform the metric into the spherically symmetric slices. First, we use (\ref{35}) to get
\begin{eqnarray}
\sinh^2\left(\delta_b\tilde{b}\right)=\frac{\gamma^2\delta_b^2\left(b_0^2X+X+2b_0\right)}{\left(b_0+X\right)^2}X\,,\label{38}\\
\gamma^2-\frac{\sinh^2\left(\delta_b\tilde{b}\right)}{\delta_b^2}=\gamma^2b_0^2\frac{1-X^2}{\left(b_0+X\right)^2}\,,\label{39}
\end{eqnarray}
where $X\equiv\tanh\left(b_0T/2\right)$. Using equations (\ref{38}) and (\ref{39}), the metric function $-g_{xx}=\frac{\tilde{p}_b^2}{\tilde{p}_cL_0^2}$ can be written as
\begin{equation}
-g_{xx}=\frac{4m^2\left(b_0^2X+X+2b_0\right)(b_0+X)^2X}{\tilde{p}_cb_0^4\left(1-X^2\right)^2}\label{gxx}\,.
\end{equation}
When $T\rightarrow\infty$, we have $X\rightarrow1$ and $\tilde{p}_c\rightarrow 4m^2e^{2T}$, which corresponds to the asymptotic region of the exterior spacetime. Note that constant $T$ surfaces in the exterior spacetime are timelike surfaces. The asymptotic expression of $-g_{xx}$ reads
\begin{equation}
-g_{xx}\approx\frac{\left(b_0+1\right)^4}{16b_0^4}e^{2T\left(b_0-1\right)}.
\end{equation}
It can be seen that the $g_{xx}$ would diverge when $T\rightarrow\infty$ as long as the quantum parameter $\delta_b$ is not zero (i.e., when $b_0\ne1$), no matter how carefully one fine-tunes it. The behaviour of the metric function $g_{xx}$ is shown in Figure~\ref{f2}.

On the other hand, the metric function $g_{TT}$ is given by 
\begin{equation}
g_{TT}=-\tilde{N}^2=\frac{\tilde{p}_c(b_0+X)^2}{\left(b_0^2X+X+2b_0\right)X}\,,
\end{equation}
where (\ref{38}) is used in the last equality. When $T\rightarrow\infty$, it can be approximated as $g_{TT}\approx\tilde{p}_c\approx4m^2e^{2T}$. Therefore $g_{TT}$ also diverges asymptotically. Note that the divergence of $g_{TT}$ at large $T$ can be removed by redefining the radial coordinate hence is not harmful.

If we do the transformation $r\equiv|\tilde{p}_c|^{1/2}$ and change $x\rightarrow t$ with a proper rescaling, the effective metric in the asymptotic region ($r\rightarrow\infty$) reads
\begin{equation}
ds^2|_{r\rightarrow\infty} = -r^{2(b_0-1)} d t^2 + d r^2 + r^2d\Omega^2\,.\label{12me}
\end{equation}
We will show explicitly that in the asymptotes, the AOS quantum black hole does not recover the Schwarzschild spacetime. In fact, the AOS quantum black hole is not asymptotically flat unless the quantum correction is absent ($b_0=1$), i.e., to obtain the proper classical limit in the asymptotic region, as is required of any physical model, we end up with the classical solution everywhere!

\subsection{Asymptotic non-flatness}
In order to explicitly demonstrate the asymptotic non-flatness of the AOS spacetime, we recall the definition of the so-called (weakly) asymptotically simple spacetimes. Essentially, a spacetime on a manifold $\mathcal{M}$ defined by a metric $g_{\mu\nu}$ is asymptotically simple if there exists a conformal compactification $\tilde{\mathcal{M}}=\mathcal{M}\cup\partial\mathcal{M}$ such that \cite{Wald,Andersson:2018rqm}
\begin{enumerate}
\item its metric $\tilde{g}_{\mu\nu}$ is conformal to the original metric $g_{\mu\nu}$, which can be written as $\tilde{g}_{\mu\nu}=\Omega^2 g_{\mu\nu}$,
\item every null geodesic in $\mathcal{M}$ has future and past endpoints on $\partial\mathcal{M}$,
\item the conformal factor satisfies the following conditions: 1) $\Omega>0$ on $\mathcal{M}$, 2) $\Omega=0$ and $\nabla_\alpha\Omega\ne0$ on $\partial\mathcal{M}$.
\end{enumerate}
In order to include spacetimes which may contain singularities, one can define the so-called weakly asymptotically simple manifold $\mathcal{N}$ which contains an open set $U$, such that the open set $U$ is isometric to a neighborhood of the boundary of another compactified asymptotically simple manifold. Finally, one defines an asymptotically flat spacetime if the spacetime is weakly asymptotically simple and asymptotically empty in the sense that the Ricci tensor vanishes in a neighborhood of $\partial\mathcal{M}$.

In order to recast the asymptotic metric (\ref{12me}) as a form applicable to the criteria mentioned above, the only possibility is to consider the conformal factor $\omega =r^{-2(b_0-1)}$ to remove the $r$ dependence in the $g_{tt}$ component. In this regard, it can be seen that the metric (\ref{12me}) is conformally related to the following line element:
\begin{equation}
d\tilde{s}^2=-dt^2 + d\tilde{r}^2 + \left(2-b_0\right)^2\tilde{r}^2 d\Omega^2\,,\label{bvas}
\end{equation}
where $b_0\ne1$. Therefore, the metric (\ref{12me}) is conformally related to a ``quasi-asymptotically flat" metric, which has been proven to be asymptotically simple but \textit{not} asymptotically empty \cite{Wald,Andersson:2018rqm,Nucamendi:1996ac}. In fact, the metric (\ref{bvas}) is the asymptotic spacetime of the global monopole proposed by Vilenkin and Barriola \cite{Barriola:1989hx}, which contains a deficit solid angle. Given that the metric (\ref{bvas}) is not fully asymptotically flat, the AOS metric which is related to (\ref{bvas}) via a conformal transformation, cannot be asymptotically flat either. This lets us unambiguously conclude that the metric (\ref{12me}) is not asymptotically flat.

\subsection{Curvature fall-off}
An additional clue that implies the inconsistent asymptotic behavior of the AOS black hole can be seen from the fall-off behavior of the curvature invariants when $r$ becomes large. The fall-off behavior of curvature corresponds to the physical requirement that the gravitational effects generated by a black hole should decrease when moving away from the black hole. Although the curvature invariants approach zero as $r\rightarrow\infty$, their fall-off behavior deviates significantly from those of the Schwarzschild solution. If one calculates the Kretschmann scalar $K$ of the effective metric and expand it in terms of $1/r$, then one finds
\begin{equation}
K=\frac{c_1}{r^4}+\left(\textrm{higher order of }\frac{1}{r}\right)\,,
\end{equation}
where $c_1=3\gamma^4\delta_b^4+\left(\textrm{higher order of }\gamma\delta_b\right)$ and it vanishes \textit{iff} the quantum parameter $\delta_b=0$. This shows that the asymptotic behavior of the metric is obviously not like the Schwarzschild spacetime, where $K\propto 1/r^6$. In fact, this inconsistency not only happens for the Kretschmann scalar, but also for other curvature invariants constructed solely from the Riemann tensor. Even though the quantum parameter is very tiny, the deviation of the AOS metric from that of the Schwarzschild would be greatly amplified according to the different asymptotic behaviors of their metrics. For a viable quantum corrected black hole solution, one would expect that the quantum effects occur only deep inside the black hole and not at large areal radius, which, if occurs, would indeed be problematic. No matter how small the quantum parameter is, the deviation of the AOS solution from that of the classical one would inevitably become sizable at large $r$. 

Finally, there is one more evidence to demonstrate that the AOS spacetime is not asymptotically flat. If the spacetime is asymptotically flat, which approximately  means that the spacetime reduces to the Minkowskian spacetime, the asymptotic regions should be maximally symmetric and should satisfy the following relation
\begin{equation}
R_{\alpha\beta\mu\nu}=\frac{R}{d(d-1)}\left(g_{\alpha\mu}g_{\beta\nu}-g_{\alpha\nu}g_{\beta\mu}\right)\,,\label{maximallsym}
\end{equation}
where $d=4$ is the spacetime dimension. From the asymptotic metric (\ref{12me}), we find that 
\begin{equation}
R=-\frac{2b_0\left(b_0-1\right)}{r^2}\,,\qquad R_{t\theta\theta t}=-\left(b_0-1\right)r^{2b_0-2}\,.
\end{equation}
If $b_0\ne1$, it can be immediately seen that (\ref{maximallsym}) is not satisfied when $r\rightarrow\infty$. Therefore, the metric (\ref{12me}) is not asymptotically flat.

\section{Asymptotic structure at null infinity}
After this paper appeared on arXiv, the authors of Refs.~\cite{Ashtekar:2018cay,Ashtekar:2018lag} wrote another paper \cite{Ashtekar:2020ckv}, mentioning that the asymptotic behavior of the curvature fall-off of the AOS model does not match to that of the Schwarzschild black hole, as we have mentioned in the previous section (the v2 version on arXiv of this paper arXiv:1902.07874v2). Although the authors of \cite{Ashtekar:2020ckv} gave a coordinate transformation on the metric and claimed that the asymptotic metric of the AOS model approaches to the Minkowski metric $\eta_{\mu\nu}$ as $1/r$, we would like to argue that such a coordinate transformation is problematic, so is the asymptotic structure of the so-derived metric. More explicitly, we will show in this section that if the asymptotic metric given in Ref.~\cite{Ashtekar:2020ckv} is correct, the existence of the quantum parameter will inevitably destroy the asymptotic structure of the spacetime.

According to Ref.~\cite{Ashtekar:2020ckv} (see Eqs.~(4.11) and (4.12) in that paper), under the following coordinate transformation
\begin{equation}
\tau=t\left(\frac{r}{2m}\right)^\epsilon\,,
\end{equation} 
the asymptotic metric of the AOS spacetime can be written as
\begin{eqnarray}
g_{\mu\nu}dx^\mu dx^\nu&=&\left(-d\tau^2+dr^2+r^2d\Omega^2\right)+\left(\frac{2m}{r}\right)^{1+\epsilon}d\tau^2\nonumber\\&+&2\epsilon\frac{\tau}{r}\left[1-\left(\frac{2m}{r}\right)^{1+\epsilon}\right]drd\tau\nonumber\\&-&\left[\left(1-\frac{1}{1-\left(\frac{2m}{r}\right)^{1+\epsilon}}\right)-\epsilon^2\frac{\tau^2}{r^2}\left(1-\left(\frac{2m}{r}\right)^{1+\epsilon}\right)\right]dr^2\,,
\end{eqnarray}
where $\epsilon\equiv b_0-1$. Although this metric is written in the form of $\eta_{\mu\nu}+\mathcal{O}(1/r)$, it cannot describe properly the asymptotic spacetime. For example, considering a \textit{would-be} null-like surface $\tau\pm r=\textrm{const.}$ and $d\Omega=0$, one can see that this surface is in fact not a null-like surface at the asymptotic region because
\begin{equation}
g_{\mu\nu}dx^\mu dx^\nu=\left(2\epsilon+\epsilon^2\right)dr^2+\mathcal{O}\left(\frac{1}{r}\right)\,,
\end{equation}
which does not vanish in the presence of the quantum parameter $\epsilon=b_0-1$. Therefore, the existence of the quantum parameters in the AOS model does destroy the asymptotic structure of the spacetime, which is not supposed to happen for a local gravitational object.

\section{Conclusions}
The recently proposed AOS effective black hole \cite{Ashtekar:2018cay,Ashtekar:2018lag} is constructed within the framework of LQG in order to resolve the classical singularity problem of the Schwarzschild spacetime. This model has several intriguing physical outcomes. In the interior region, the Schwarzschild singularity is indeed replaced with a spacelike transition surface connecting a trapped and an anti-trapped region. This transition surface results from the quantum corrections introduced in the theory, and these quantum corrections quickly decrease when moving away from the transition surface. For a macroscopic black hole, the curvature scale acquires its maximum value at the transition surface and this maximum value turns out to be independent of the black hole mass.  The authors extended their solution to the exterior region. It was argued \cite{Ashtekar:2018cay,Ashtekar:2018lag} that such exterior spacetime is asymptotically flat and a corresponding ADM mass can be defined. Based on the ADM mass defined in this model, it was argued that there is no mass amplification when crossing the transition surface. It is in fact rather exciting that the AOS model is able to address so many long-lasting issues haunting in the LQG community, and this model definitely deserves a deeper scrutiny.

In this letter, we have explicitly shown that the AOS exterior region is actually not asymptotically flat. In addition, even though the curvature invariants approach zero in the asymptotic region, their fall-off behaviors do not recover those of the Schwarzschild spacetime. This means that the AOS model \textit{does not} reduce to the Schwarzschild black hole in the asymptotic region, at parametrically large areal radius. In addition, we have found that after a proper conformal transformation, the transformed AOS effective metric shares the same asymptotic expression of the global monopole spacetime proposed by Vilenkin and Barriola \cite{Barriola:1989hx}, which is asymptotically simple, but not asymptotically flat. The AOS spacetime being asymptotically non-flat can also be seen from the viewpoint of how the curvature tensors behave in the asymptotic region.

Even though it is possible to define an appropriate ADM mass in some asymptotically non-flat spacetimes, such as the global monopole spacetime \cite{Nucamendi:1996ac}, it is actually not clear whether this can be done for the AOS spacetime. In any case, the AOS quantum black hole does not reduce to the Schwarzschild black hole when $r\rightarrow\infty$ in the presence of the quantum parameters. In fact, after we had pointed out this issue in \cite{Bouhmadi-Lopez:2019hpp}\footnote{Here we refer specifically to the v1 version of the paper, which was posted on arXiv on 21 Feb 2019.}, the author of \cite{Bojowald:2019dry} has proven that the AOS exterior spacetime violates general covariance. The effective Hamiltonian can be expanded in terms of the quantum parameters and it has been shown that the first order terms in quantum parameters violate general covariance. Therefore, the exterior spacetime of AOS black hole being asymptotically non-flat might be a manifestation of its violation of general covariance. Thus, in the presence of quantum parameters, the AOS model leads to unphysical conclusions, which lead us to conclude that the singularity resolution in this model comes at a very heavy price of spoiling its classical limit, at least in what concerns the asymptotic property of the spacetime. 

\vspace{3mm}

Note added: Although the authors of \cite{Ashtekar:2020ckv} agree with the fact that the curvature fall-off is slower compared to Schwarzschild spacetime, they claim that this is not harmful because the deviation is extremely small if the quantum parameters are small. The authors also provide a coordinate transformation showing that the metric can be recast into a Minkowskian one in the asymptotic region (See Eqs.~(4.11) and (4.12) in \cite{Ashtekar:2020ckv}). However, as we emphasize in this paper, in the asymptotic region, any deviation compared with the classical solution resulting from quantum corrections is a manifestation of a pathology of the effective spacetime, no matter how small it is. In addition, the coordinate transformation (4.11) in \cite{Ashtekar:2020ckv} is extremely problematic to examine the asymptotic limit, i.e., when $r \rightarrow \infty$. It is so because, in this limit $r\rightarrow\infty$, the new coordinate  $\tau$ is finite \textit{only when} $t\rightarrow 0$. Keeping in mind that $\tau$ appears explicitly in some of the metric components (see Eqn. (4.12) in \cite{Ashtekar:2020ckv}), it is clear that such a coordinate transformation breaks down when $t$ is large  and cannot be used to probe the asymptotic limit of the AOS model. Furthermore, the asymptotic metric (4.13) is not valid at null infinity, as we have shown explicitly in the previous section. This can be proven by showing that $\tau\pm r=\textrm{const.}$ is not a null-like surface in the metric (4.12). More explicitly, considering the \textit{would-be} null-like surface $\tau-r=\textrm{const.}$ and $d\omega=0$ in Eqn.~(4.12), and making a series expansion in terms of $1/r$, the leading order of the right-hand side of Eqn.~(4.12) becomes $\left(2\epsilon+\epsilon^2\right)dr^2$, which is nonzero asymptotically. Therefore, the presence of the quantum parameter $\epsilon\equiv b_0-1$ does destroy the asymptotic structure of the spacetime.

\ack{MBL is supported by the Basque Foundation of Science Ikerbasque. She also would like to acknowledge the partial support from the Basque government Grant No. IT956-16 (Spain) and from the project FIS2017-85076-P (MINECO/AEI/FEDER, UE). CYC and PC are supported by Ministry of Science and Technology (MOST), Taiwan, through No. 107-2119-M-002-005 and No. 108-2811-M-002-682, Leung Center for Cosmology and Particle Astrophysics (LeCosPA) of National Taiwan University, and Taiwan National Center for Theoretical Sciences (NCTS). PC is in addition supported by US Department of Energy under Contract No. DE-AC03-76SF00515. The research of SB and DY is supported in part by the Ministry of Science, ICT \& Future Planning, Gyeongsangbuk-do and Pohang City and the National Research Foundation of Korea grant no. 2018R1D1A1B07049126. SB is also supported in part by funds from NSERC, from the Canada Research Chair program and by a McGill Space Institute fellowship.}

\end{document}